\documentstyle[preprint,aps]{revtex}
\def\braket#1#2{\left\langle{#1}\right\vert\left.{#2}\right\rangle}
\def\ket#1{\left\vert{#1}\right\rangle}
\def\bra#1{\left\langle{#1}\right\vert}
\def\bracket#1#2#3{\left\langle{#1}\left\vert{#2}\right\vert{#3}\right\rangle}

\def\nint{\int\nolimits}
\def\ds{\displaystyle}
\renewcommand{\dag}{\dagger}
\newcommand{\D}{ \mbox{$\cal{D}$} }

\begin{document}
\draft
\title{ Semiclassical Quantization for the Motion of Guiding Center 
Using the Coherent State Path Integral}
\author{T.Tochishita, M.Mizui and H.Kuratsuji}
\address{Department of Physics, Ritsumeikan University-BKC,
Kusatsu City 525-77, JAPAN}
%
\date{\today}
\maketitle
%
\vspace*{-1.0truecm}
\begin{abstract}
\begin{center}
\parbox{14cm}{
A new form of the Bohr-Sommerfeld quantization is given of
the motion of guiding center in strong magnetic field.
This is obtained by the effective action 
for the degree of guiding center which is deduced from 
the coherent state path integral for the two types of motion 
under the mutual interaction; the cyclotron motion and 
the motion of guiding center.
}
\end{center}
\end{abstract}
%
%
\narrowtext

\section{INTRODUCTION}

The physics of charged particles in strong magnetic field has
been one of the central problems in
quantum mechanics inspired by condensed matter physics
\cite{Springer}.
The unique feature is that the system is described by two degrees
of freedom, namely, 
the motion of cyclotron and guiding center, if one
chooses the special type of gauge for writing the vector potential.
In the case that we have the uniform
magnetic field only, the guiding center degrees of freedom is not active; 
this is manifested
by the infinite degeneracy of Landau levels.
The degeneracy implies some symmetry which is governed
by nothing but the guiding center coordinates. 
Now if a potential of non-magnetic origin is added, the guiding center degre
es 
of freedom 
begins to be active. The energy exchange occurs 
between the motion of cyclotron and guiding center, namely,
we expect the mixing between inter-Landau levels. Thus the problem is
reduced to the study of the two systems which couple each other. 

In this note we focus our attention on the
motion of the guiding center. We are
particularly interested in the effective action to
get the semiclassical quantization rule.
 In order to achieve this, we use
the coherent state path integral. It is well known  that both the motion of
cyclotron and guiding center can be described by the raising 
and lowering operators
for harmonic oscillator and hence the coherent state provides
with a natural means for this specific problem. \cite{Kahn}
Concerning the semiclassical quantization, Entelis and Levit studied
it in the framework of the Born-Oppenheimer approximation\cite{Levit92}
, in which the motion of guiding center and the cyclotron motion 
are treated as the slow and fast degrees of freedom respectively. 
The fast cyclotron motion is
firstly handled to get the adiabatic effective action for
the slow motion of guiding center that includes the adiabatic quantum
phase factor
\cite{Kura}.
On the contrary, our method does not rely upon the use of
the specific assumption of adiabaticity.
This paper is a preliminary account of the basic idea and the
more detailed argument will be given in the forthcoming paper.

\section{Coherent state}

We shall start with a brief account for the coherent state representing the
quantum state of the charged particle in uniform magnetic field.
Let us introduce the guiding center coordinates\cite{Kahn},\cite{Kubo}
\begin{equation}
X = x - \Pi_y /\mu\omega \quad ,\quad Y = y + \Pi_x / \mu\omega
\end{equation}
with $ \omega = {e B \over \mu c} $ being the cyclotron frequency.
$ \vec{\Pi} = \vec{p} + { e\over c } \vec{A} $ represents 
the kinematical momentum 
where the vector potential is given by
$ \vec{A} = ( -{1\over 2} B y,{1\over 2} B x, 0 ) $ .
We have the commutation relation
\begin{equation}
\left[ X, Y \right] = {i \hbar \over \mu\omega}
\end{equation}
Now we define the following boson creation and annihilation operators
\begin{equation}
\left\{
\begin{array}{ccc}
   a &=& \ds{1 \over \sqrt{2\mu\omega} } ( \Pi_x - i \Pi_y )  \\
   a^{\dag} &=& \ds{1 \over \sqrt{2\mu\omega} } ( \Pi_x + i \Pi_y )
\end{array}
\right.
\quad
\left\{
\begin{array}{ccc}
 b  &=& \ds{1 \over \sqrt{2} \ell } ( X + i Y ) \\
 b^{\dag} &=& \ds{1 \over \sqrt{2} \ell }( X - i Y )
\end{array}
\right.
\end{equation}
which satisfy the commutation relation; $ [a , a^{\dag} ] = 1 $
and $ [ b , b^{\dag} ] = 1 $ and $a$ and $b$ commute each other, 
where $\ell$ is the so-called magnetic length 
$ \ell \equiv \sqrt{\hbar\over \mu\omega}  $. 
In terms of these boson operators, 
the Hamiltonian in the uniform magnetic field 
is given by 
\begin{equation}
H_0   = \left( a^{\dag} a + {1\over 2} \right) \hbar\omega 
\end{equation}
The angular momentum operator $L_z = xp_y - yp_x $ is also written as 
\begin{equation}
L_z =( a^{\dag} a - b^{\dag} b )\hbar
\end{equation}
Now let us define $\ket{N, m }$ as
the simultaneous eigenstate of
$ H_0  $  and $ L_z $,
\begin{equation}
\begin{array}{ccc}
H_0\ket{N, m} &=& \left( N + \ds{1 \over 2} \right)\hbar\omega\ket{ N, m } \\
L_z\ket{ N, m } &=& m\hbar \ket{ N, m }
\end{array}
\end{equation}
which satisfy the following relations
\begin{equation}
\left\{
\begin{array}{ccc}
  a^{\dag} \ket{N, m} &=& ( N + 1 )^{1 \over 2}\ket{N+1, m+1} \\
  a        \ket{N, m} &=& N^{1 \over 2}\ket{N-1, m-1} \\
  b^{\dag} \ket{N, m} &=&  ( N - m + 1 )^{1 \over 2}\ket{N, m-1} \\
  b        \ket{N, m} &=&  ( N - m )^{1 \over 2}\ket{N, m+1}
\end{array}
\right.
\end{equation}
Here $N, m$ take the value such that $ N = 0, 1, 2, \cdots\cdots,
\infty, m = -\infty,\cdots,N - 1, N $. This feature indicates that
the Landau level labelled by the quantum number $ N $ has an infinite number 
of degeneracy. Thus $ \ket{N,m} $ can be constructed as follows;
\begin{equation}
\ket{N,m} = {( b^{\dagger})^{N-m}(a^{\dagger})^N \over N! (N-m)! }\ket{0,0}
\end{equation}
Now the coherent state $\ket{ z, \xi }$ is introduced such that
\begin{equation}
   a \ket{ z, \xi } = z \ket{ z, \xi },
   b \ket{ z, \xi } = \xi \ket{ z, \xi }
\end{equation}
with $ z, \xi $ represent the complex number.
$\ket{ z, \xi }$ is constructed as
\begin{equation}
   \ket{ z, \xi } = \exp \left( z a^{\dag} + \xi b^{\dag} \right) \ket{ 0,0 }
\end{equation}
Note that this forms the so-called "overcomplete set of states".
The overlap between two CS's is calculated as
\begin{equation}
\braket{ z^{\prime}, \xi ^{\prime} }{ z, \xi }
    = \exp \left[-{1 \over 2}\left(
       |z|^2 + |z^{\prime}|^2 + |\xi|^2 + |\xi^{\prime}|^2 \right)
           + (z^{\prime})^* z + (\xi ^{\prime})^*  \xi \right]
\end{equation}

The most important property of the coherent state is
the resolution of unity,
\begin{equation}
\int\ket{ z,\xi }d\mu [z,\xi]\bra{ z,\xi }  = 1
\end{equation}
and $ d\mu(z,\xi) $ means the measure of the complex parameter
space given as
\begin{equation}
d\mu[z,\xi] = {dz^r dz^i \over \pi} {d\xi^r d\xi^i \over \pi}
\end{equation}

The wave function representing the coherent state is
written as
\begin{equation}
\begin{array}{lcl}
\psi_{z, \xi}( x, y )
   &\equiv& \braket{ x, y }{ z , \xi } \\
   &=&\ds{1\over \sqrt{2\pi} \ell}
  	\exp \left\{
         - \ds{1\over 4\ell^2} \left[ x - \sqrt{2}\ell ( iz +  \xi ) \right]^2
         - \ds{1\over 4\ell^2} \left[ y + \sqrt{2}\ell ( z  + i\xi ) \right]^2
            +iz\xi \right\}
\end{array}
\end{equation}
which is used to evaluate the matrix elements of the Hamiltonian.
\section{Coherent state path integral}

We now consider the quantum transition in terms of the path
integral using the coherent state constructed above.
In the following argument
we are concerned with the one-body problem for the charged particle
in uniform magnetic field as well as the force of non-magnetic origin;
$ H = H_0 + \phi(x,y)$. 
The nonmagnetic term $ \phi(x,y) $ involves the coupling between the
guiding center and cyclotron motion. Here we make the
following remark. In order to deal with the path integral for two systems
under the mutual interaction, the two steps procedure is
adopted, namely, one first considers the
transition amplitude as if one degree were fixed, and after
having accomplished this,  the transition amplitude for the remaining
degree is considered. Now, we consider the transition amplitude
\begin{equation}
K = \bracket{\xi_f, z_f}{\exp[-{i\over \hbar}\hat H t]}{\xi_i,z_i}
\end{equation}

Following the usual procedure of the coherent state path integral,
we first divide the time $ t $ into N small interval ( and
take the limit $ N \rightarrow \infty $ ) and next
insert the resolution of unity relation for the
coherent state at each time division point, then $ K $
is cast into the infinite dimensional integral \cite{su2}
\begin{equation}
\begin{array}{lcl}
K(\xi_f,z_ft \vert \xi_i,z_i,0)
	&=	&\ds\lim_{N \rightarrow \infty}
	\int \cdots \int
		\prod^{N - 1}_{k = 1} d \mu(\xi_k) 
		\prod_{k = 1}^N
	\bracket{\xi_k}{\exp[ -\ds{i \over \hbar} \hat H \epsilon]}{\xi_{k -1}}
\end{array}
\end{equation}
The infinitesimal factor can be approximated 
in the case of the limit $ \epsilon \rightarrow 0 $;
\begin{equation}
\begin{array}{lcl}
\bracket{\xi_k}{\exp[ -\ds{i \over \hbar} \hat H \epsilon]}{\xi_{k - 1}}
	&\simeq & \bracket{\xi_k}{(1 -\ds{i \over \hbar} \hat H \epsilon)}
		{\xi_{k - 1}} \\
	&= & \braket{\xi_k}{\xi_{k - 1}} 
			\left( 1 - \ds{i \over \hbar} \epsilon
	\ds{ \bracket{\xi_k}{\hat H}{\xi_{k - 1}} \over \braket{\xi_k}
	{\xi_{k - 1}} } \right).
\end{array}
\end{equation}
The first factor is just the overlap between the infinitesimal interval,
which is explicitly written as
\begin{equation}
   \braket{\xi_k}{\xi_{k-1}} = \exp[\xi_k^* \xi_{k-1}
        -{\vert \xi_k \vert^2 \over 2} - {\vert \xi_{k-1} \vert^2 \over 2}]
\label{aho}
\end{equation}
By using this form, the propagator turns out to be
\begin{equation}
\label{x}
 K(\xi_f, z_f,t: \xi_i,z_i,0) = \int \exp[{i\over \hbar}\int
     {i\hbar \over 2}(\xi^{*}\dot \xi - c.c)dt]T(z_f,z_i)
     \D \mu[\xi(t), \xi^{*}(t)]
\end{equation}
$ T(z_f, z_i) $ means the amplitude for the transition
from the initial state $ \ket{z_i} $ to the final state $  \ket{z_f} $
under the motion of the guiding center $ \xi(t) $, namely,
\begin{equation}
T(z_f, z_i)= \bracket{z_f}{T\exp[-{i\over \hbar}
   \nint^T_0 h(a^{\dagger}, a; \xi(t), \xi^{*}(t))dt]}{z_i}
\label{sakata}
\end{equation}
Here $ h(a^{\dagger}, a; \xi, \xi^{*})
\equiv \bracket{\xi}{H(a^{\dagger}, a, b^{\dagger}, b)}{\xi} $,
which implies that in calculating the transition amplitude
for the cyclotron motion the degree of guiding center is replaced
as if it were the c-number coordinate. Thus, if using the relation
of the resolution of unity holding for $ \ket{z} $ ,
$ T(z_f, z_i) $ is
written in terms of the path integral over the $ z(t) $ space
\begin{equation}
T(z_f,z_i) = \int \exp[{i\over \hbar}\nint^T_0
       {i\hbar \over 2}\{(z^{*}\dot z - c.c)
           - (z^{*}z + {1\over 2})\hbar\omega
                - \bracket{z}{h}{z}\}dt]\D\mu(z(t))
\end{equation}
where the expectation value $ \bracket{z}{h}{z} $ is given by
\begin{equation}
\Phi(z,z^*,\xi,\xi^*)
     = \bracket{z}{h}{z} \\
     = {\bracket{z,\xi}{\phi ( x,y )}{ z,\xi}\over
              \braket{z,\xi}{z,\xi}}
\end{equation}

\section{Semiclassical Quantization Formula}

We shall consider the semiclassical limit of the
propagator given by the form (\ref{x}). The semiclassical limit
means that $ \hbar \rightarrow 0 $.
Noting that the propagator (\ref{aho}) consists of integral over
two different degrees of freedom, we take the semiclassical limit in two-step
procedure, namely, in the first step, the semiclassical limit is
taken for the integration for the cyclotron degree,
thus
\begin{equation}
   T(z_f, z_i) = \exp[{i\over \hbar}\Gamma(\tilde C)]
\exp[-{i\over \hbar}\nint^T_0
             \bracket{z}{\hat h(\xi(t),\xi^{*}(t))}{z}dt]
\label{g}
\end{equation}
where $ \Gamma(\tilde{C}) $
\begin{equation}
    \Gamma(\tilde{C}) = \int \bracket{z}{i\hbar{\partial
\over {\partial t}}}{z}dt
\label{ga}
\end{equation}
and  $ \tilde C $ represents the orbit satisfying
the action principle
\begin{equation}
   \delta\nint^T_0 \bracket{z}{i\hbar{\partial \over
{\partial t}} -\hat h(\xi(t), \xi^{*}(t))}{z}dt = 0.
\label{h}
\end{equation}

The variation equation yields the equation of motion for the complex
variables $ z, z^{*} $, which leads to the classical equation of motion
in the coupled form
\begin{equation}
\left\{
\begin{array}{lcc}
i\hbar\dot{z}^{*} + \hbar\omega z^{*}
	+ \ds{ \partial \Phi \over \partial  z } 	= 0  \\
-i\hbar \dot{z} + \hbar\omega z
	+ \ds{ \partial \Phi \over \partial  z^{*}} 	= 0  \\
\end{array}
\right.
\end{equation}
Note that the solution of these equations
are given in terms of the functions of the orbit $ \xi(t) $.
Thus the evolution operator $ K $ becomes
\begin{equation}
    K_{eff}(T) = \int \exp[{i\over \hbar}S_{eff}]\prod d\mu[C]
\label{i}
\end{equation}
where the effective action is given by
\begin{equation}
      S_{eff}  =  S_0 + \hbar\Gamma
            - \nint^T_0 \bracket{z}{\hat h}{z}dt
\label{j}
\end{equation}
with $ S_0 = \int { i\hbar \over 2} (\xi^{\dagger}\dot{\xi}  -c.c)dt. $
If we note that $ z$ and $ z^{*} $ are given by
the functions of the orbit $ \xi $, the canonical term
is written as
\begin{equation}
   \bracket{z}{i\hbar{\partial
\over {\partial t}}}{z} = \left\{\bracket{z}
{i\hbar{\partial \over {\partial z}}}{z}{{\partial z} \over {\partial \xi}}
      + \bracket{z}{i\hbar{\partial \over {\partial z^{*}}}}{z}
       {{\partial z^{*}} \over {\partial \xi}}\right\}{d\xi \over dt}
\label{l}
\end{equation}
The canonical term in the effective action for the motion of guiding
center is thus changed by an amount of $ \omega $, where
\begin{equation}
    \omega = (A_z^i + A_{\bar z}^i)d\xi_i .
    \label{m}
\end{equation}

Here $ A $ represents the so-called "connection field" that
is defined as
\begin{equation}
\begin{array}{lcl}
A_z^i 	& = & \bracket{z}{i\hbar\ds{\partial \over \partial z}}{z}
		\ds{{\partial z} \over {\partial x_i}} \\
A_{\bar z}^i
	& = &  \bracket{z}{i\hbar\ds{\partial \over {\partial z^{*}}}}{z}
\ds{{\partial z^{*}} \over {\partial x_i}}
\end{array}
\end{equation}
where $ x_i $ denotes the abbreviation of the guiding center
coordinate (X,Y).
The appearance of the phase $ \Gamma $ arises from taking
the semiclassical limit of the expression of the time-ordered
product in (\ref{sakata}). This process is a counterpart of
the adiabatic process which picks up a specific path from an
infinite sum given by the time-ordered product.
\cite{Kura}

Now if we use the effective action of the form (\ref{j}), 
we can obtain the semiclassical quantization formula for
the motion of guiding center.
We first consider the semiclassical approximation
for the effective propagator; $ K_{eff}^{sc}(T) $.
 Having carried out it, 
 we next take the Fourier transform of $ K_{eff}^{sc}(T) $
 to apply the stationary phase approximation over the
 $ T $-integral, and finally taking into account of the contribution of
multiple traversals of the basic periodic oribit, we have
\begin{equation}
K_{sc}(E) = \sum {\exp[iW(E)] \over 1 - \exp[iW(E)]}
\end{equation}
with $ W(E) = \oint {i\hbar \over 2}(\xi^{*}\dot \xi -c.c)dt + \Gamma(C) $.
Then, the quantization rule should follow from poles on the E-plane
\begin{equation}
    \oint {i\hbar \over 2}(\xi^{*}\dot \xi -c.c)dt =
          (n-{\Gamma(C) \over 2\pi})2\pi\hbar
\label{y}
\end{equation}
which just gives the corrected Bohr-Sommerfeld formula including
the geometric phase $ \Gamma $. $ C $ represents the periodic
orbits determined by the action principle $ \delta S_{eff} = 0 $,
which turns out to be
\begin{equation}
\left\{
\begin{array}{lcc}
i\hbar\dot{\xi}^{*}
	+ \ds{ \partial \Phi \over \partial \xi } 	&=& 0  \\
-i\hbar\dot{\xi}
	+ \ds{ \partial \Phi \over \partial \xi^{*} } &=& 0  \\
\end{array}
\right.
\end{equation}
The formula  (\ref{y}) is just a generalization of the previously
result that is the one including the effect of the adiabatic
phase.
\cite{Levit92}

\section{Simple example}

We consider the case of the central force;  $
\phi ( x,y ) = \phi (r), r = \sqrt{x^2 + y^2} $
for which we get $ \Phi ( z,z^*,\xi,\xi^* ) = \Phi ( |\chi|^2 ) $		where  $ 
\chi $
and hence the equation of motion yields
\begin{equation}
\left\{
\begin{array}{ccc}
i\hbar\dot{z}^{*} + \hbar\omega z^{*} + \ds{i\over 2} (-i z^{*} + \xi )
\ds{d \Phi \over d |\chi|^2}&=& 0  \\
-i\hbar \dot{z} + \hbar\omega z - \ds{i\over 2} (i z + \xi^{*} )
\ds{d \Phi \over d |\chi|^2}  	&=& 0 \\
i\hbar\dot{\xi}^{*} + \ds{1\over 2} ( iz + \xi^{*} ) 
			\ds{d \Phi \over d |\chi|^2}
		&=& 0 \\
-i\hbar\dot{\xi} + \ds{1\over 2} ( - iz^{*} + \xi ) 
			\ds{d \Phi \over d |\chi|^2}
		&=& 0
\end{array}
\right.
\end{equation}
As a special case of the lowest Landau level; $N = 0 $
namely, $ z = z^* = 0 $, the Lagrangian turns out to be
\begin{equation}
L  = {i\hbar\over 2} ( \xi^* \dot{\xi} - \dot{\xi}^* \xi )
  -  {1\over 2}  \hbar\omega - \Phi ( |\xi|^2 )
\end{equation}
which yields the equation of motion
\begin{equation}
\left\{
\begin{array}{ccc}
i\hbar\dot{\xi}^{*} + \ds{1\over 2} \xi^{*} \ds{d \Phi \over d |\xi|^2}
		&=& 0 \\
-i\hbar\dot{\xi}    + \ds{1\over 2} \xi \ds{d \Phi \over d |\xi|^2}
		&=& 0
\end{array}
\right.
\end{equation}
from which we get $ {d (|\xi|^2) \over dt} = 0  $, 
thus $|\xi|^2 = {\rm const} $.
Physically speaking, the restriction to
the lowest Landau level means the strong magnetic
field limit $B \longrightarrow \infty$ or in terms of the
cyclotron frequency, $ \omega \longrightarrow \infty
~~( \omega \propto B ) $.
Now we shall consider the Bohr-Sommerfeld quantization,
which is obtained as special case of (\ref{y}), namely,
by putting $ z= z^{*} = 0 $,
\begin{equation} {i \hbar\over 2} \oint_C ( \xi^* d \xi - \xi d \xi^* )
=  2\pi m \hbar \end{equation}
If we put $ \xi = {1 \over \sqrt{2} \ell } ( X + i Y ) ,
 \xi^* = {1 \over \sqrt{2} \ell } ( X - i Y ) $,
this can be rewritten as the form
\begin{equation}
\begin{array}{lcl}
\ds{i \hbar\over 2} \oint_C \ds{i \over l^2} ( X dY - Y dX )
	&=& - \ds{ \hbar \over 2 \ell^2 } \int_S 2 dX \wedge dY \\
	&=& - 2\pi \hbar|\xi|^2
\end{array}
\end{equation}
Thus, we have
$$ |\xi|^2 = - m
$$
and hence the energy spectrum is obtained as
\begin{equation} E_m = {1\over 2}\hbar\omega - \Phi ( - m )
\end{equation}
Note that the above quantization rule gives nothing but the
angular  momentum quantization , since $ \vert \xi \vert^2 $
is given as the expectation value of $ L_z $ with respect to
the coherent state.

As a special case, we consider the Coulomb potential
for the interaction  $ \phi (r) = - {\Large { e^2 \over r }}$,
for which $ \Phi(\vert \xi \vert ) $
becomes
\begin{equation}
\Phi(\vert \chi \vert ) =
{  e^2 \pi\over\sqrt{2\pi} l}
 			{\rm e}^{-{|\chi|^2/2} } I_0 ( |\chi|^2/2 )
\end{equation}
where $ I_0(\vert \xi \vert) $ stands for the modified Bessel
function. The energy spectrum is obtained by substituting the relation
$ \vert \xi \vert^2 = -m $.

\section{Final Remarks}

In this note we have developed a basic idea for the effective action
for the motion of guiding center to yield the semiclassical quantization rule.
The formula presented here is mainly concerned with the single
particle in strong magnetic field together with the potential
of non-magnetic origin. Here an outline is given of the possible
problems that are expected to be developed on basis of the present
 general formalism. (i): To take into
account of the cyclotron motion in the quantization rule,
which is to consider the effect of the geometric phase $ \Gamma $
after all. (ii): To extend the case of the
lowest Landau level to the many particle case, which may be
regarded as a similarity with the vortex gas for which the
force is given by the logarithmic potential
\cite{Kuratsuji92}.
(iii): In relation to 
the many-particle extension, it may be interesting to consider the
charged particle with two kind of charges, namely, with
opposite charges each other,  which may be analogous to the
two kind of vortices with opposite vortex strength.
These will be left as subjects for the future study.

\end{document}